# Local Structure Controlling the Glass Transition in a Prototype Metal-Metalloid Glass

Jason J. Maldonis, Paul M. Voyles
Department of Materials Science and Engineering, University of Wisconsin-Madison, Madison, WI, USA

Abstract

We use a structure analysis technique that quantifies cluster geometries using machine learning to identify a bicapped square antiprism (BSAP) as the nearest-neighbor cluster involved in the glass transition in $Pd_{82}Si_{18}$, a prototype metal-metalloid metallic glass. BSAPs have slow dynamics and grow significantly in concentration during cooling, like icosahedra in other systems. These results are evidence that some preferred structures universally contribute to the glass transition and show that analysis of interatomic geometry can find important structures not revealed by topology.

Topological descriptors have had great success as abstract descriptions of the structure of disordered materials. Topological descriptors are often derived from a connectivity graph with atoms as vertices and atomic bonds as edges. The connectivity graph preserves the topology of the structure but abandons the atomic coordinates. Various topological descriptors are derived from the connectivity graph, which differ in the structural length scale they describe. For example, common neighbor analysis (CNA), which is highly localized, produces a triplet of integers that describes the topology of two atoms and their joint neighboring atoms [1]. Voronoi index (VI) analyses [2] are less localized than CNA and describe the topology of an atom and all of its nearest neighbors, typically 9-16 atoms in metallic glass systems. Rings statistics in covalent glasses involve 2[nd] and more distant neighbors [3].

VI analysis is perhaps the most common abstraction of metallic glass atomic structure. It is derived from the Voronoi tessellation of a material's structure, which describes the structure as the set of 3D polyhedra surrounding every atom and consisting of the space closer to that atom than any other atom. The tessellation is a complete geometrical representation of a structure with a one-to-one correspondence to the atomic positions. The representation becomes topological when the polyhedra are abstracted into a set of integers $<n_3, n_4, n_5, n_6, ...>$ that report the number of faces of the polyhedra with $n_i$ edges.

The VI representation of metallic glass structure has significant explanatory power in some systems. For example, in metal-metal MGs, atoms with icosahedral topology such as <0 0 12 0> were shown to correlate with properties including glass-forming ability [4–7], dynamic heterogeneities [8,9], atomic mobility [5,10], and deformation behavior [11,12]. In particular, in Cu-Zr based MGs [13], icosahedral clusters are slow-moving [5,10] and the fraction of icosahedral clusters increases dramatically as the supercooled liquid approaches the glass transition [4–7]. It is hypothesized that the glass transition is related to this dramatic increase in the fraction of slow-moving clusters. An increase in the population of slow-moving clusters results in the propagation of rigidity through the liquid and, at a critical threshold, causes the substantial increase in viscosity that defines the glass transition [14–18].

VI analysis is less useful for metal-metalloid glasses. The dominant topology from VI analysis is the tricapped trigonal prisms, but their population does not change significantly with temperature, nor are they reported to exhibit slow dynamics [4,19,20]. So the question remains: Do local structures exist in metal-metalloid glasses that contribute to large increases in viscosity at the glass transition in the same manner that icosahedra do in metal-metal systems?

Here we employ a recently developed structure analysis technique called *motif extraction* [21,22] to characterize the structure of a prototype metal-metalloid glass, $Pd_{82}Si_{18}$. Motif extraction identifies a bicapped square antiprism (BSAP) structure which has slow dynamics and increases in concentration as the system cools through the glass transition. These results show evidence in simulations for universality of slow, preferred structures playing a central role in the glass transition of metallic glasses and show that geometric structural analysis may succeed where topological analysis does not.

Motif extraction [21] applies machine learning clustering techniques to local structures composed of an atom and its nearest neighbors (henceforth called *clusters*). The machine learning algorithm identifies collections of similar clusters, and then the similar clusters' structures are averaged to create a single structure (a *motif*) that is representative of all the clusters in the collection. The set of motifs represent the statistically significant local structures in the disordered material, and each cluster in the glass can be represented abstractly by the motif to which it is most similar. Motif extraction uses geometry rather than topology to describe the local structure of glasses and therefore preserves information about the relative atom positions.

We applied motif extraction to a $Pd_{82}Si_{18}$ MG with 9826 atoms quenched using molecular dynamics (MD) in an NPT ensemble from the liquid state at 1800 K to 0 K at $5\times10^{10}$ K/s in LAMMPS [23,24] with an embedded atom potential [19] and a Nose-Hoover thermostat and barostat (at 0 atm). We extracted atomic configurations every 50 K from 1800 K to 0 K and calculated their inherent structures by conjugate gradient energy minimization in LAMMPS using the same potential. All analyses were performed on the inherent structures except propensity of motion, which was calculated from MD snapshots. We calculated partial structure factors, $S(q)$, using RINGS [25] and the VI of each atom using Voro++ [26]. The alignment algorithm [27] used by motif extraction was used to assign each atom to the motif most similar to it. Propensity for motion [28,29] was calculated from the distance each atom moved after 40 ps in an NPT ensemble at 1200 K, averaged over 100 model trajectories initialized with random Maxwell-Boltzmann velocity distributions [28]. For the propensity of motion calculations, the VI of each atom was calculated from the MD snapshot configuration rather than from the inherent structure because the snapshot configuration, and trajectories generated from it, contains the information about the dynamics of the atoms. The dynamics information is lost upon performing the conjugate gradient minimization that produces the inherent structure.

Figure S1 shows the internal energy and volume of the system during the MD quench. The simulated $T_g$ is 744 K, estimated from the change in volume of the system. Figure S2 shows consistency between the simulated X-ray partial structure factors and previous simulation results on the same system [19]. In the glassy state the nearest neighbor coordination numbers (CNs) range from 8-15 (see Figure S3). Si has CN 10 or lower and Pd has CN 11 or higher.

Figure 1(a) and (b) show the evolution of the VI and CN distribution in the models as a function of temperature during cooling. The CN distribution in Figure 1(a) changes significantly, especially CN 11, which decreases with cooling through $T_g$, and CN 13, which increases. This behavior is in strong contrast to a typical metal-metal glass, $Zr_{50}Cu_{45}Al_5$, for which the CN distribution is largely temperature independent (see Figure S4). The VI distribution in Figure 1(b) has significant temperature dependence, but most of the changes are a result of the changes in CN, not a result of atomic rearrangements at constant CN (as in metal-metal glasses). Figure 1(c) shows the fraction of each VI divided by the fraction of the corresponding CN at each temperature, which normalizes out the effects of the changing CN on the VI distribution. In Figure 1(d), the curves are offset so that the high-temperature concentration falls near zero, which highlights the changes in the fraction of CN of each VI. This data captures changes in local topology that are not created by the changing CN distribution. In particular, clusters with VI <0 2 8 0> (CN 10) are more prevalent in the glass than in the liquid, although this change is smaller than the change in icosahedral order in metal-metal glasses [4–7,13].

Neither the change in topology associated with the change in CN nor the change in topology at constant CN identifies the clusters associated with dynamic arrest at $T_g$. Figure 2 shows the propensity of motion at 1200 K of the 25 most common VIs in the $Pd_{82}Si_{18}$ liquid, sorted from slowest to fastest. The clusters with VI <0 2 8 0>, which increase in population, are not particularly slow-moving, unlike icosahedral topologies in metal-metal glasses [5,10]. The other topology that dominates a particular CN, <0 3 6 0> for CN 9, is slower than most of the other topologies, but its concentration changes little during cooling (see Figure 1(d)). <0 2 8 2> and <0 1 10 2> topologies have the lowest propensity for motion, but neither one is particularly prevalent, and their fraction of CN changes by less than 5%. In contrast to metal-metal systems, icosahedral topologies are not slow moving, nor do they change significantly in population through $T_g$. Unfortunately, VI analysis does not help us understand the glass transition in this material.

Figure S5 shows the 27 unique motifs identified using motif extraction. We use the notation $n_A$ to label each motif where $n$ is the CN of the motif and $\{A, B, C, \ldots\}$ enumerates the motifs at constant $n$. Table S1 lists the motifs, CN, VI, and dissimilarity [27] to the topologically close-packed Z-cluster with the same CN (see Ref [21] for details). As in a previous analysis of $Zr_{50}Cu_{45}Al_5$ [21], a hierarchy of motifs are identified with increasing coordination number. Some motifs have striking symmetry, such as motif $14_A$ and the motifs (denoted with superscript $Z$) are very similar to Frank-Kasper polyhedra or Z-clusters [13]. Some CNs have only one motif, while others have several. The atomic coordinates of the motifs are provided in the Supplementary Information.

Motif extraction reveals a strong correlation of structure with the glass transition. Out of the 27 unique motifs identified, $10_A^Z$ and $10_B$ are of particular interest (shown in Figure 3(a,b)). These motifs have the same VI, <0 2 8 0>, and CN 10, but motif extraction decoupled them. The fraction of CN 10 atoms most similar to motif $10_A^Z$ increases dramatically through the glass transition, and the fraction of motif $10_B$ decreases dramatically (see Figure 4(a)). These opposing changes largely cancel each other out in the fraction of CN 10 clusters with VI <0 2 8 0>. In addition, the propensity for motion of $10_A^Z$ clusters is amongst the lowest in the system (see Figure 4(b)), and the propensity for motion of $10_B$ clusters is higher than all but one other structure. The three motifs slower than $10_A^Z$, motifs $15_A^Z$, $14_B$, and $9_Z$, change very little in concentration through $T_g$. The motif faster than $10_B$, motif $8_A$, occurs at very low density in the glass.

Motif $10_A^Z$ is therefore a slow-moving structure whose population increases as the liquid approaches the glass transition. It is geometrically similar to the BSAP shown in Figure 3(c)), which is the most close-packed polytetrahedral structure with CN 10 [21]. The structure can be described as one center atom aligned vertically with an atom on the top and another on the bottom, and two rings of four atoms in horizontal planes above and below the center atom, rotationally offset by 45°. An icosahedron is similar, except that the four-atom rings are replaced with five-atom rings and the rotational offset is 36°. However, unlike an icosahedron, a BSAP has a crystallographically allowed rotational symmetry. A BSAP has four-fold symmetry, but only about one axis, shown in Figure 3(c), whereas an icosahedron has six axes of five-fold symmetry.

Although we have focused on BSAP clusters in $Pd_{82}Si_{18}$ here, there are icosahedral clusters in $Pd_{82}Si_{18}$ and BSAP clusters in $Zr_{50}Cu_{45}Al_5$ [21]. In $Pd_{82}Si_{18}$, there is a motif with VI <0 0 12 0>, $12_C$ (Figure 3(d)), but geometrically, it is barely icosahedral compared to the <0 0 12 0> motif in $Zr_{50}Cu_{45}Al_5$ (Figure 3(e)), which has nearly perfect icosahedral geometry. The $12_C$ motif represents 5% of the glassy model and has intermediate propensity for motion (Figure 4(b)). Given the unfavorable atomic radii and the potential for some influence of covalent bonding [30], the lack of icosahedral structure in $Pd_{82}Si_{18}$ is not surprising. In $Zr_{50}Cu_{45}Al_5$, the population of BSAP clusters increased significantly as the supercooled liquid approached $T_g$, with only the icosahedral motif increasing more dramatically.

Overall, these results support the hypothesis that there are preferred nearest-neighbor atomic clusters that exist in the liquid state and both grow in concentration and slow in dynamics more quickly than other structures as the liquid is cooled through the glass transition [31–33]. This phenomena is reminiscent of the cooperative rearranging regions and growing dynamical length scale of physical theories of the glass transition [34,35], although proving that connection is beyond the time scales accessible to molecular dynamics simulations like ones analyzed here. The nature of that local structure of course depends on the materials system. For metals, our results suggest that a small set of motifs are responsible for the glass transition, with icosahedra dominant in some glasses and BSAPs dominant in others. That said, we cannot rule out that in systems with different bonding characteristics, both icosahedral and BSAP geometries could be unfavorable and a different motif might play the same functional role.

Finally, these results demonstrate the utility of a structural abstraction involving geometry, as opposed to topology, in identifying structure-property correlations. Whether the BSAP structure is connected to other properties like plastic deformation the same way icosahedra are remains to be seen, but the properties of a material are fundamentally controlled by interatomic energies, and these energies are strongly determined by relative atom positions, *i.e.* geometry. Because topological analysis techniques discard information about the relative atom positions, the information contained in topological descriptions of structure is less directly linked to the atomic energies. Therefore, while motif extraction and VI analysis provide complementary information, we believe (and our evidence shows) that motif extraction provides a structural description that is more closely connected to materials' properties.

In conclusion, we have identified a close-packed, local structure motif, the bicapped square antiprism, in $Pd_{82}Si_{18}$, a model metal-metalloid glass-forming alloy, that plays a significant role in the glass transition. This motif is among the slowest moving local structures in the supercooled liquid, and it increases in concentration significantly through the glass transition. This motif was identified using *motif extraction*, a machine learning method for extracting the geometry of characteristic structure from disordered atomic models. The BSAP motif has the same VI (<0 2 8 0>) as another local structure with the opposite behavior, which masks the structure-property relationship from topological measures like VI. These results support the general conclusion that across metallic liquids, preferred nearest-neighbor atom clusters exist which play an outsized role in dynamic arrest during the glass transition.

This work was supported by the U. S. National Science Foundation DMREF program (DMR-1728933). Computational resources were provided by the UW-Madison Center For High Throughput Computing, which is supported by UW-Madison, the Advanced Computing Initiative, the Wisconsin Alumni Research Foundation, the Wisconsin Institutes for Discovery, and the National Science Foundation.

**Figures**

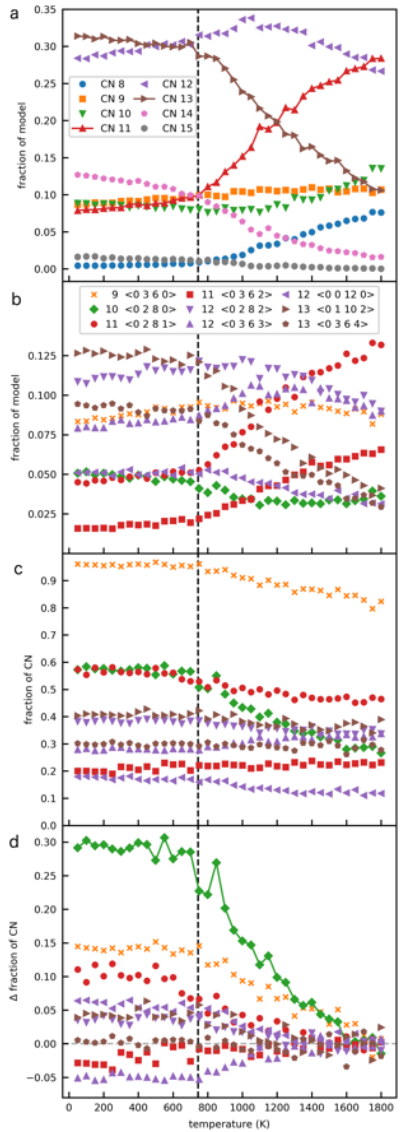

Figure 1: (a) The fractions of clusters with different CNs in a $Pd_{82}Si_{18}$ MG change significantly with temperature, in contrast to a $Zr_{50}Cu_{45}Al_5$ glass (see Figure S4). (b) The fraction of clusters with different VI also change as a function of temperature. (c) Normalizing the fraction of clusters by the fraction of CN shows that changes in CN drive most changes in topology. (d) Replotting the data in (c) with each curve offset to zero at high temperature emphasizes the changes in topology through the glass transition. The fraction of clusters with topology <0 2 8 0> shows a significant increase during cooling. The fraction per CN of clusters with VI <0 0 12 0> is low.

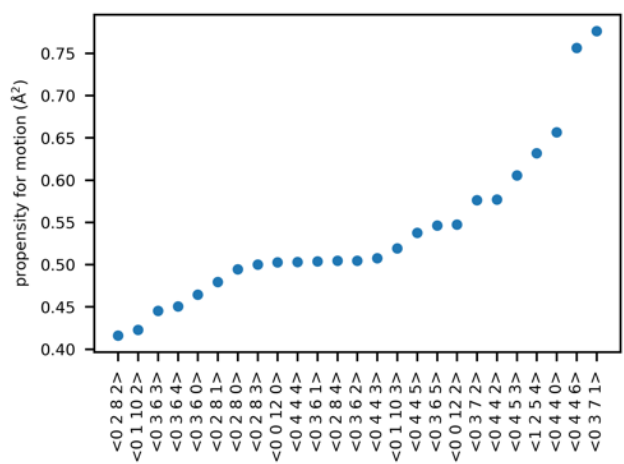

Figure 2: The propensity for motion at 1200 K of the 25 most common VI categories do not reveal an immediate connection between structure and properties. In particular, clusters with VI <0 2 8 0> have intermediate propensity for motion.

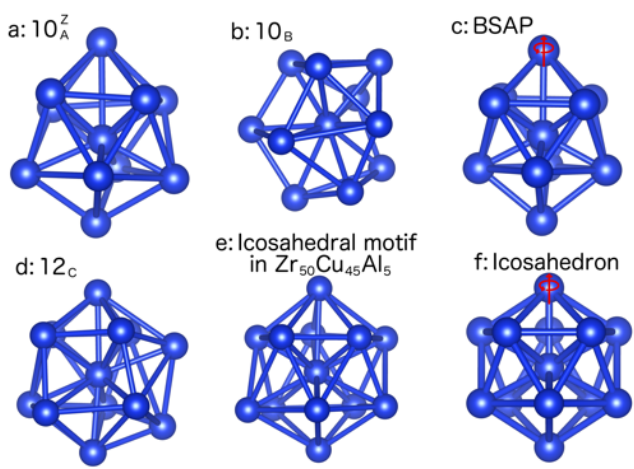

Figure 3: Motifs in $Pd_{82}Si_{18}$ compared to other structures. (a) Motif $10_A^Z$, (b) motif $10_B$, (c) a perfect BSAP, the Z-cluster with CN 10 [13,21], (d) motif $12_C$, the most icosahedral motif in $Pd_{82}Si_{18}$, (e) the most icosahedral motif in $Zr_{45}Cu_{45}Al_5$ [21] and (f) a perfect icosahedron. $10_A^Z$ is similar to the perfect BSAP, but $10_B$ is not. $12_C$ is significantly less geometrically icosahedral than the extracted motif for $Zr_{50}Cu_{45}Al_5$ in (e), as illustrated by the square face in the foreground. The BSAP in (c) has a single 4-fold rotation axis shown by the red arrow. One of the 5-fold rotation axes of the icosahedron is shown in (f), but a similar axis exists about every straight line of atoms passing through the center atom.

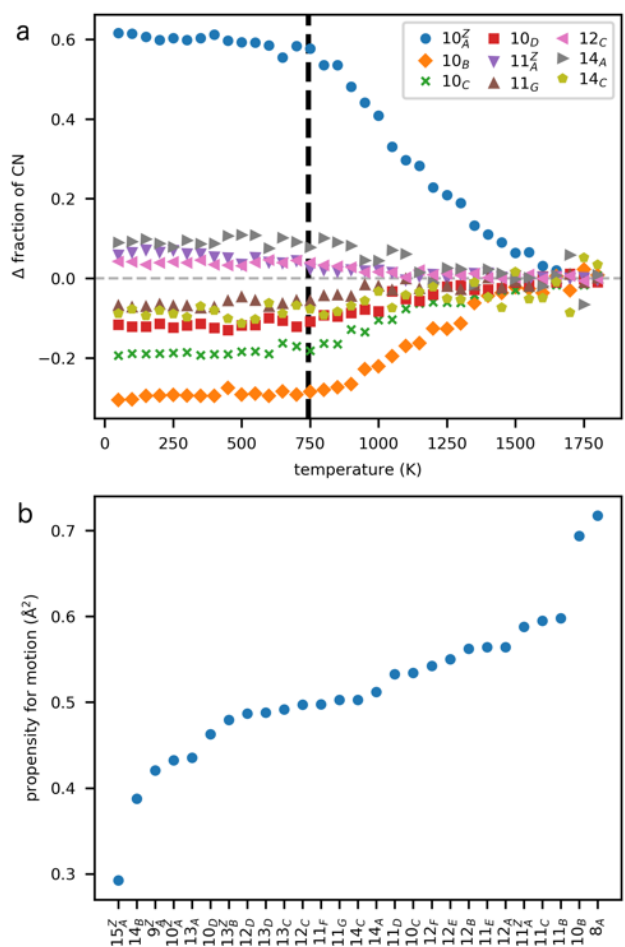

Figure 4: (a) The change in the fraction of CN for the eight motifs whose fraction of CN changes the most during cooling and motif $12_C$, normalized as in Figure 1(d). (b) The propensity for motion at 1200 K of each motif. $10_A^Z$ is amongst the slowest motifs, and $10_B$ is amongst the fastest. In general, close-packed Z clusters tend to have a low propensity for motion.

**Supplemental Information**

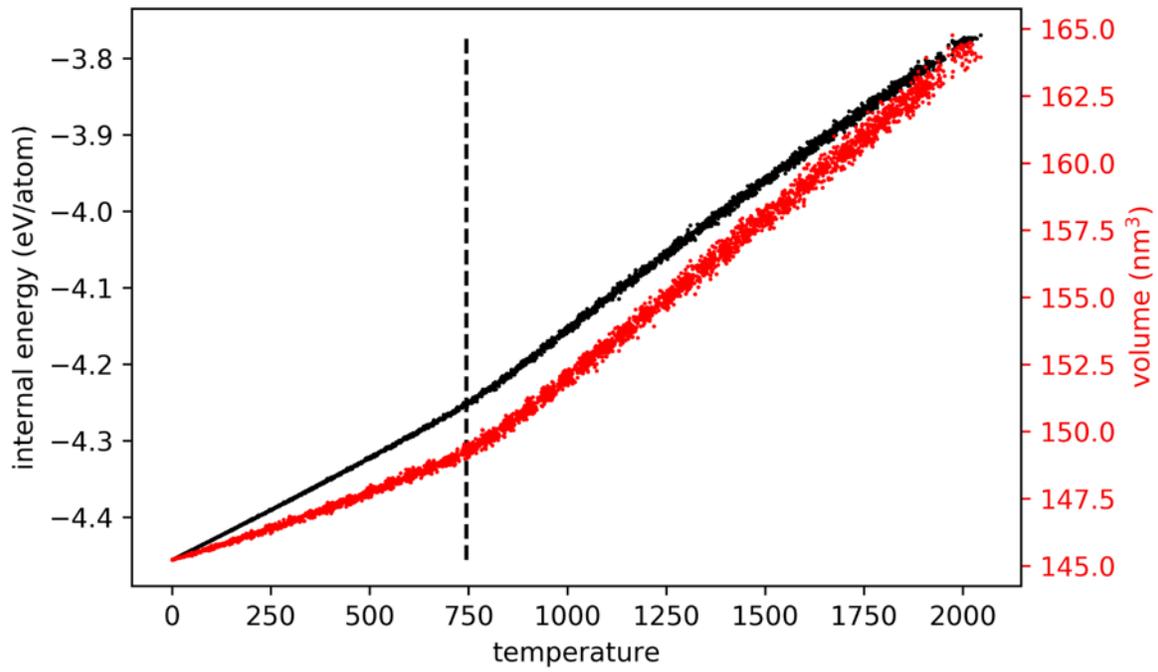

Figure S1: Internal energy (black) and volume (red) vs. temperature. The dashed black line shows $T_g$ = 744 K.

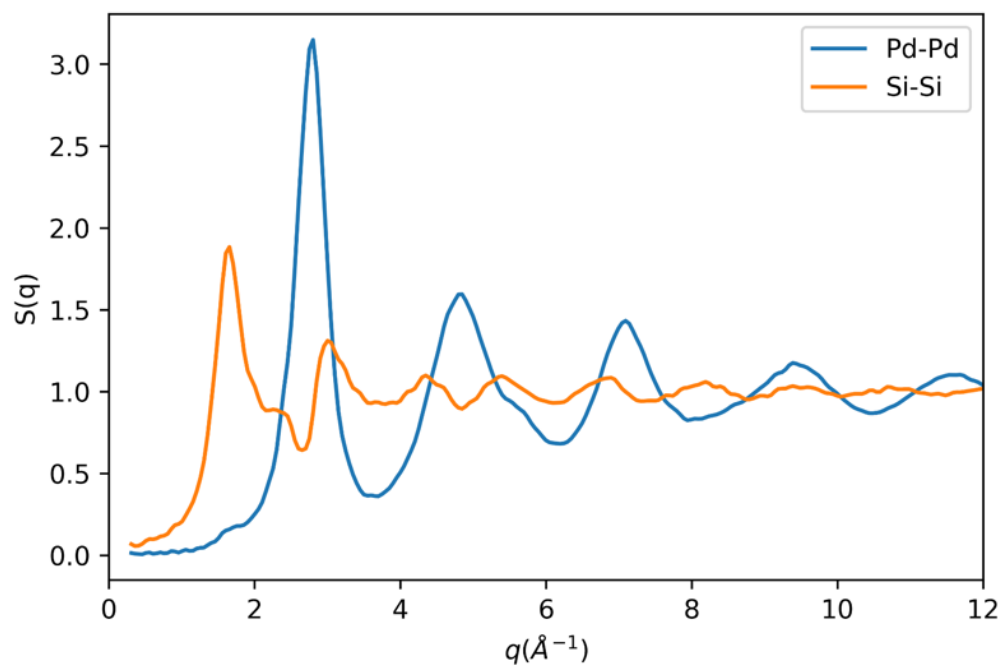

Figure S2: X-ray partial structure factors for the $Pd_{82}Si_{18}$ liquid at 1200K agree well with previous results for the same potential [19].

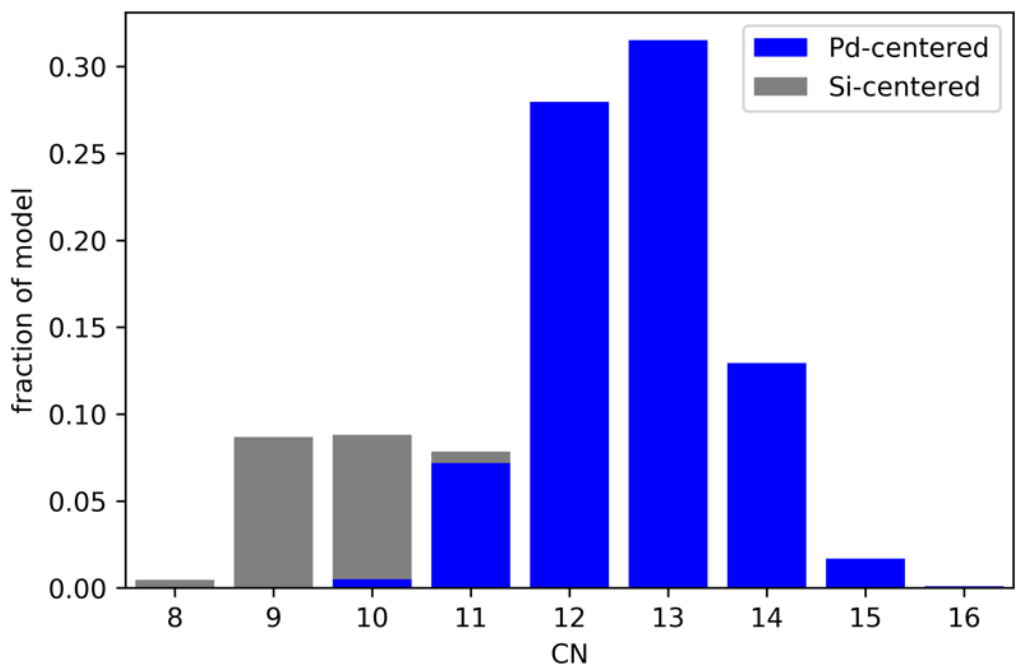

Figure S3: (a) CN distribution at 300 K colorized by atomic specie. Nearly all Si atoms are 9- or 10-coordinated, while Pd atoms tend to be 11-14 coordinated.

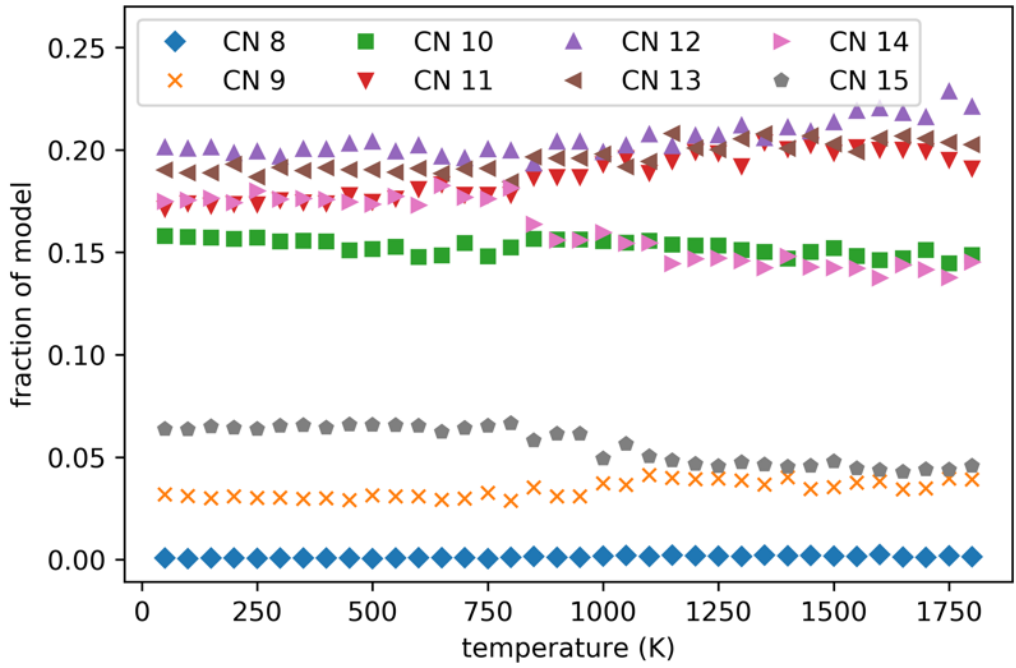

Figure S4: The CN distribution as a function of temperature in a $Zr_{50}Cu_{45}Al_5$ MG, see Ref. [21]. Contrasting this data with the data in Figure 1(a), the changes in CN with temperature are much more significant in $Pd_{82}Si_{18}$.

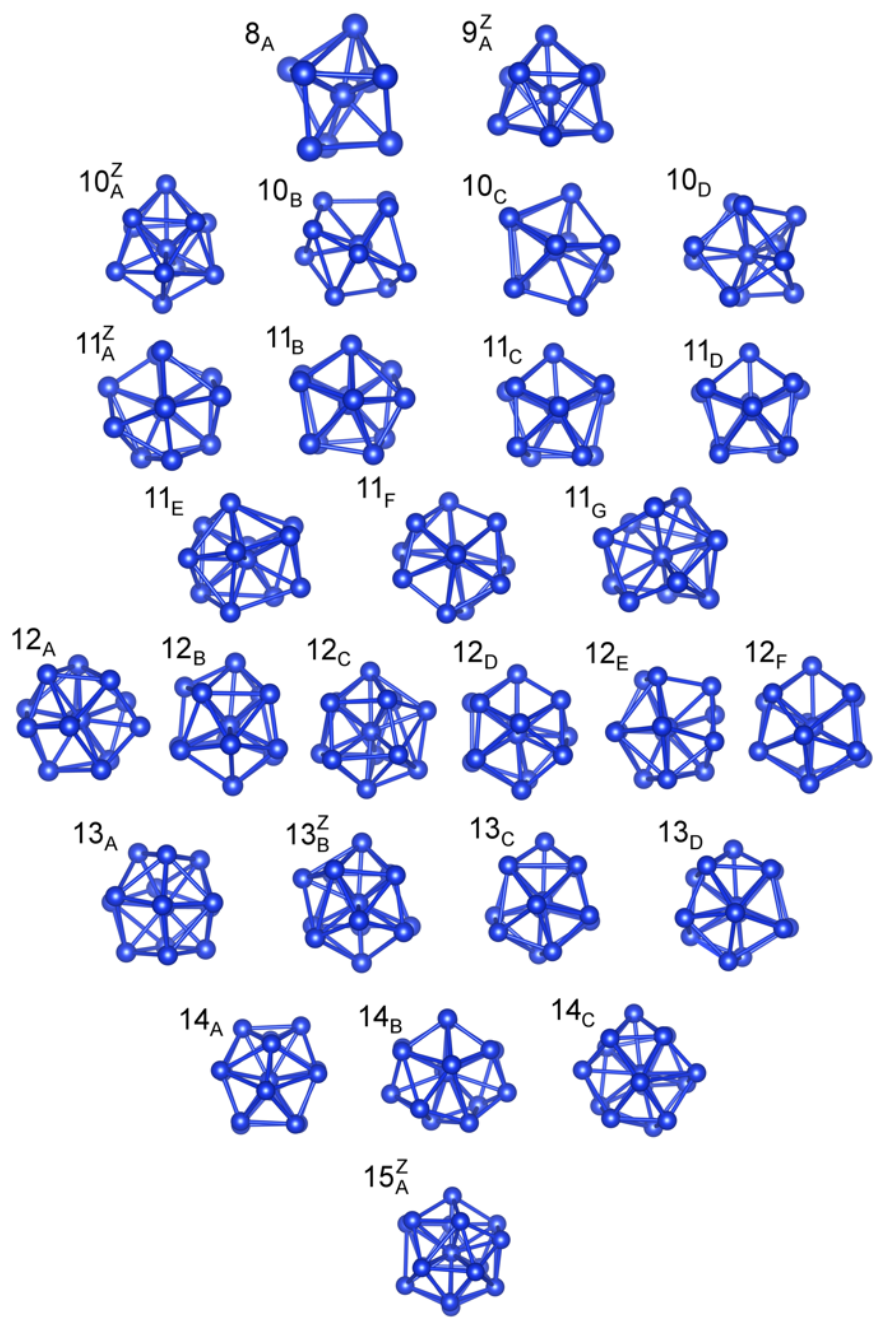

Figure S5: The 27 motifs identified in a $Pd_{82}Si_{18}$ MG arranged by CN. Orientations were chosen to illustrate various symmetry elements, if any exist. Atomic coordinates for these clusters may be found in the SI.

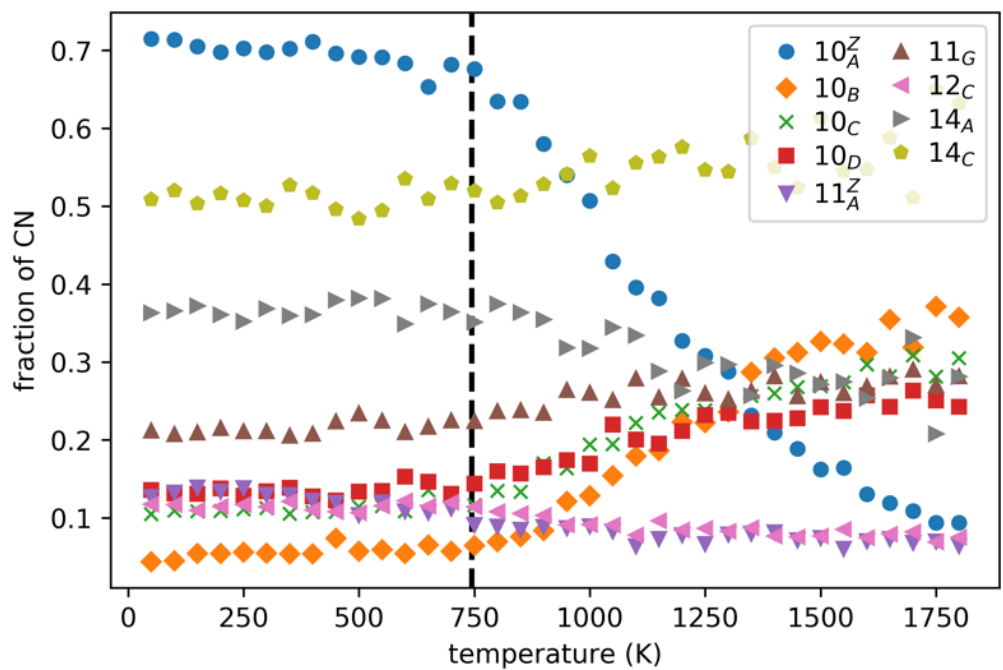

Figure S6: Data from Figure 4(a) without offsetting the curves to 0.0 at high temperatures. This data shows both the dramatic increase in the fraction per CN of motif $10_A^Z$ as well as the low fraction per CN of motif $12_C$.

Table S1: The CN, VI, and dissimilarity ($D$) [27] to the Z-cluster with the same CN for all motifs. For all the pairs of clusters in the model, dissimilarities range from 0.0 to 1.97 with a mean of 1.0 and a standard deviation of 0.16.

| Motif Label | CN | VI | Dissimilarity ($D$) to Z-cluster |
|---|---|---|---|
| $8_A$ | 8 | <0 4 4 0> | 0.718 |
| $9_A^Z$ | 9 | <0 3 6 0> | 0.636 |
| $10_A^Z$ | 10 | <0 2 8 0> | 0.675 |
| $10_B$ | 10 | <0 2 8 0> | 1.165 |
| $10_C$ | 10 | <0 4 4 2> | 1.101 |
| $10_D$ | 10 | <0 3 6 1> | 1.003 |
| $11_A^Z$ | 11 | <0 2 8 1> | 0.606 |
| $11_B$ | 11 | <0 2 8 1> | 0.781 |
| $11_C$ | 11 | <0 2 8 1> | 0.999 |
| $11_D$ | 11 | <0 3 6 2> | 0.939 |
| $11_E$ | 11 | <0 2 8 1> | 0.828 |
| $11_F$ | 11 | <0 2 8 1> | 0.874 |
| $11_G$ | 11 | <0 2 8 1> | 0.872 |
| $12_A$ | 12 | <0 0 12 0> | 0.984 |
| $12_B$ | 12 | <0 3 6 3> | 1.339 |
| $12_C$ | 12 | <0 0 12 0> | 0.860 |
| $12_D$ | 12 | <0 3 6 3> | 1.125 |
| $12_E$ | 12 | <0 2 8 2> | 1.142 |
| $12_F$ | 12 | <0 2 8 2> | 1.052 |
| $13_A$ | 13 | <0 1 10 2> | 0.684 |
| $13_B^Z$ | 13 | <0 1 10 2> | 0.386 |
| $13_C$ | 13 | <0 2 8 3> | 0.938 |
| $13_D$ | 13 | <0 1 10 2> | 0.576 |
| $14_A$ | 14 | <0 2 8 4> | 0.768 |
| $14_B$ | 14 | <0 2 8 4> | 0.971 |
| $14_C$ | 14 | <0 1 10 3> | 0.947 |
| $15_E^Z$ | 15 | <0 0 12 3> | 0.432 |
| Z8 | 8 | <0 4 4 0> | 0. |
| Z9 | 9 | <0 3 6 0> | 0. |
| Z10 | 10 | <0 2 8 0> | 0. |
| Z11 | 11 | <0 2 8 1> | 0. |
| Z12 | 12 | <0 0 12 0> | 0. |
| Z13 | 13 | <0 1 10 2> | 0. |
| Z14 | 14 | <0 0 12 2> | 0. |
| Z15 | 15 | <0 0 12 3> | 0. |